\documentclass[aps,pra,twocolumn,superscriptaddress]{revtex4-1}
\usepackage{graphicx}
\usepackage{MnSymbol}
\usepackage{bm}  % bold math

\newcommand  {\at}    {\mathrm{at}}
\newcommand  {\sub}   {\mathrm{sub}}
\newcommand  {\supp}  {\mathrm{sup}}
\newcommand  {\scat}  {\mathrm{sc}}
\newcommand  {\Imag}  {\mathrm{Im}}
\newcommand  {\Real}  {\mathrm{Re}}
\newcommand  {\tr}    {\mathrm{tr}}
\newcommand  {\LL}    {\mathrm{LL}}

\newcommand  {\Ein}   {\mathrm{Ein}}
\newcommand  {\diff}  {\mathrm{diff}}
\newcommand  {\NF}    {\mathrm{NF}}
\newcommand  {\FF}    {\mathrm{FF}}

\newcommand {\INLN} {Universit\'e C\^ote d'Azur, CNRS, INLN, France}
\newcommand {\Constantine} {Laboratoire de Physique Mathematique et Subatomique, Universit\'e des Fr\`eres Mentouri,
Constantine, Algeria}

\begin{document}

\title{Light interacting with atomic ensembles: collective, cooperative and mesoscopic effects.}

\author{W. Guerin}
\email{william.guerin@inln.cnrs.fr}
\affiliation{\INLN}
\author{M. T. Rouabah}
\affiliation{\INLN}
\affiliation{\Constantine}
\author{R. Kaiser}
\affiliation{\INLN}

\date{\today}

\begin{abstract}
Cooperative scattering has been the subject of intense research in the last years.
In this article, we discuss the concept of cooperative scattering from a broad perspective. We briefly review the various collective effects that occur when light interacts with an ensemble of atoms. We show that some effects that have been recently discussed in the context of `single-photon superradiance', or cooperative scattering in the linear-optics regime, can also be explained by `standard optics', i.e., using macroscopic quantities such as the susceptibility or the diffusion coefficient.
We explain why some collective effects depend on the atomic density, and others on the optical depth. In particular, we show that, for a large and dilute atomic sample driven by a far-detuned laser, the decay of the fluorescence, which exhibits superradiant and subradiant dynamics, depends only on the on-resonance optical depth.
We also discuss the link between concepts that are independently studied in the quantum-optics community and in the mesoscopic-physics community. We show that the coupled-dipole model predicts a departure from Ohm's law for the diffuse light, that incoherent multiple scattering can induce a saturation of fluorescence and we also show the similarity between the weak-localization correction to the diffusion coefficient and the inaccuracy of Lorentz local field correction to the susceptibility.
\end{abstract}

\maketitle

\section{Introduction}

`Cooperative scattering' of light by a collection of two-level atoms has become a hot topic in the recent years, with a lot of experimental results on superradiance~\cite{Goban:2015,Bromley:2016,Araujo:2016,Roof:2016}, subradiance~\cite{Guerin:2016} and `cooperative' shifts~\cite{Roehlsberger:2010,Keaveney:2012,Okaba:2014,Meir:2014,Jenkins:2016,Roof:2016} in the `single-photon superradiance' regime~\cite{Scully:2006,Scully:2009,Scully:2009b}, or equivalently in the linear-optics regime. By contrast, the experiments performed in the 1970s-1980s were performed in a regime where a large number of atoms are excited~\cite{Feld:1980,Gross:1982}. Related experiments with multi-level atoms~\cite{Oliveira:2014} or artificial atoms~\cite{Tighineanu:2016} illustrate that the phenomena studied in two-level systems can find their application in other fields of research. Despite this recent interest, it seems that a common agreement on what should be called cooperativity is still lacking. It has recently been suggested that `cooperativity' should be reserved to effects that could not be explained by `the mean-field approach of traditional optics'~\cite{Javanainen:2016}, and that such effects would appear for dense atomic samples only~\cite{Javanainen:talk}. Here, `dense' means that the atomic density $\rho$ should not be very small compared to $k^3$, where $\lambda =  2\pi/k$ is the wavelength of the atomic transition. However, our recent results on sub- and superradiance~\cite{Dicke:1954} show that these effects are controlled by the resonant optical depth of the sample and not by its density~\cite{Guerin:2016,Araujo:2016}. If one agrees that Dicke sub- and superradiance in extended samples~\cite{Arecchi:1970,Rehler:1971} deserve being called `cooperative' effects, the restriction to high spatial densities should clearly be dropped.

% Rest unchanged

Although everyone certainly agrees on the physics, terminology confusion does not help in understanding this topic. It can also be a real barrier when trying to communicate with experts from other, neighboring, fields. With this problematic in mind, in this article, we would like, first, to discuss the concept of `cooperative scattering', and, second, to point out the possible link between the so-called `mean-field approach of standard optics' used in~\cite{Javanainen:2016} and what is called the `independent scattering' approximation in mesoscopy~\cite{Lagendijk:1996,Rossum:1999,AkkermansMontambaux}.

Unfortunately, terminology confusion has always been present in this field. For instance, Bonifacio and Lugiato proposed the term `superfluorescence' instead of superradiance when there is no initial macroscopic dipole in the atomic system but only uncorrelated excited atoms~\cite{Bonifacio:1975}. As soon as propagation effects are involved, superfluorescence can also be seen as the transient regime of `amplified spontaneous emission'~\cite{Feld:1980}, a well-known, and somewhat trivial phenomenon in laser physics, although a difference does exist and has been studied in Ref.~\cite{Malcuit:1987}. Dicke himself spoke of a mirrorless `coherence brightened laser'~\cite{Dicke:1964}. Another example is the analogy or confusion between subradiance~\cite{Pavolini:1985,Guerin:2016} and `radiation trapping'~\cite{Holstein:1947,Molisch,Labeyrie:2003}. The difference and early confusion between the two phenomena was pointed out by Cummings~\cite{Cummings:1986}.
%Finally, the huge amount of literature on the subject (Dicke's paper~\cite{Dicke:1954} has been cited more than 3700 times) makes almost impossible to get a comprehensive overview. It also induces the risk of independently discovering something that was already published many years earlier and was largely forgotten.

Although we do not pretend to be able to solve all controversies, we feel that it may be instructive to tackle these questions from a broader perspective, which could be, in general, the study of all \emph{collective} effects in light-atom interactions, or the transition from microscopic properties of light scattering by atoms to macroscopic, optical and transport properties of extended and/or dense atomic vapors. This subject is even older than Dicke's paper, as it was investigated by Rayleigh, Lorentz, Lorenz, Ewald, Oseen, and many others~\cite{BornWolf}.

%\section{Collective effects in light-atom interactions}

We will focus on experiments in which near-resonant light from a laser is exciting a sample of $N$ two-level atoms at rest. The experimental observables that can be measured include extensive quantities that increase with the atom number, such as the total amount of scattered light, or some intensive quantity that could \textit{a priori} be independent of the atom number. If an extensive quantity is not simply given by the atom number times the same quantity for a single-atom, or if an intensive quantity does depend on the atom number, nontrivial collective effects probably are at work.
A first example of an extensive quantity is the total amount of scattered light, which, \emph{in some limit}, is proportional to the atom number. Another example of an extensive quantity is the dephasing accumulated by the wave crossing the sample, which is also increasing with the atom number, and the notion of refractive index, which is a macroscopic (collective) quantity provides an efficient description. Collective effects thus take place in most situations, with some effects depending on the atomic density $\rho$, and others on the size $L$ of the medium, as detailed in the following.
Note that as soon as propagation effects or quantities associated to transient phenomena become relevant, the size of the sample will play an important role, and the important parameter turns out to be the optical depth. On the other side, for quantities that can be defined in steady state for a bulk medium, the density is the relevant parameter.

\section{Collective effects that depend on the density}

Obviously, all effects or quantities that can be defined for an infinite medium depend on the atomic density $\rho$, which is the only parameter characterizing the medium.

\subsection{At low atomic density}

The most simple example is the medium susceptibility $\chi$. At low density, it is related to the atomic polarizability $\alpha$ by $\chi = \rho \alpha$. For simple two-level atoms and in the linear-optics regime (low intensity), the polarizability is
\begin{equation}
\alpha = \frac{6\pi}{k^3} \times \frac{-1}{2\Delta/\Gamma+i} = \frac{6\pi}{k^3} \times \frac{-2\Delta/\Gamma + i}{1+4\Delta^2/\Gamma^2} \, .
\end{equation}
The complex refractive index $n$ is defined from the susceptibility by $n^2 = 1+\chi$. At low density, $\chi \ll 1$ and $n \simeq 1+\chi/2$. Inside the medium, the wavevector is changed from $k$ to $nk$. The real part of the refractive index is thus responsible for the dispersive properties of the material and its imaginary part for the attenuation of the wave. One can thus define a characteristic extinction length (or linear extinction coefficient) for the intensity of the light propagating in the incident optical mode,
\begin{equation}
\ell_\mathrm{ex}^{-1} = \rho k \Imag(\alpha) = \frac{\rho \sigma_0}{1+4\Delta^2/\Gamma^2} \, ,
\end{equation}
where $\sigma_0 = 6\pi/k^2 = 3\lambda^2/(2\pi)$ is the resonant scattering cross-section.

%Note that the polarizability can be changed by nonlinear effects of by the addition of other external fields (or lasers) such that its imaginary part can become negative. In that case, it corresponds to amplification of light.

The susceptibility is not sufficient to describe the behavior of the light inside the medium. In particular, it does not make a difference if the attenuation is due to absorption, when the electromagnetic energy is transferred to the medium, or to scattering, when the energy is just removed from the incident electromagnetic mode to feed other spatial modes. Scattering is often seen as being due to impurity, or granularity of matter, and as such, it is neglected in the susceptibility, which corresponds to a continuous-medium approximation. It is often stated, in particular in the recent literature about cooperative scattering (see, e.g., Refs.~\cite{Svidzinsky:2008,Svidzinsky:2008b,Prasad:2010}), that the continuous-medium approximation is only valid when the density is high, $\rho k^{-3} \gtrsim 1$.
%We think that this is not true. What the continuous-medium approximation does is neglecting the scattered light, as if it were absorbed, but this can be a very good approximation, even for a dilute sample, when one is interested in the coherent transmission, and not in the scattered light.
However, the continuous-medium approximation does actually neglect the scattered light and treats it as if it were absorbed. This can indeed be a very good approximation when one is interested in the coherent transmission and not in the scattered light. It is important to note that this approximation is not restricted to high spatial densities (as in condensed matter or very high density vapors) but also holds in the dilute limit where the interatomic distances are larger than the optical wavelength.
The refractive index is thus very useful for describing light propagation in dilute atomic clouds. It has been successfully used to describe subtle experiments involving nonlinear effects~\cite{Labeyrie:2014} or photonic band gaps~\cite{Schilke:2011,Schilke:2012b}. As discussed below, it turns out that the continuous-medium approximation for atomic vapors is in fact more accurate for dilute than for dense vapors~\cite{Javanainen:2016,Jennewein:2016}.
%As pointed out by Javanainen and Ruostekoski, it is actually quite the contrary: at large density, the refractive index approach encounters difficulties, as we discuss later.

In order to describe the scattered light, another approach is necessary, in which one basic quantity is the mean-free path $\ell_\scat = 1/(\rho \sigma_\scat)$. If we only consider elastic scattering, which is valid if the incoming light has a weak intensity~\cite{Cohen:PetitLivreRouge-EN}, the (elastic) scattering cross-section is related to the polarizability by~\cite{Lagendijk:1996}
\begin{equation}
\sigma_\scat = \frac{k^4}{6\pi} |\alpha|^2 \; .
\end{equation}
In a large medium of size much larger than the mean-free path, light will be scattered many times before escaping. In this case, many observables can be very well described by a diffusion equation for the electromagnetic energy density, at the condition to perform an average over the disorder configurations. This fact is often surprising for quantum opticians, who expect that light scattered off a disordered medium produces a speckle pattern. However, it is well known to people from mesoscopic physics that after averaging over the disorder, the remaining intensity distribution, or its temporal dynamics in a pulsed experiment, fits perfectly well the prediction of the diffusion equation, although this approach neglects all interference effects~\cite{footnote_RTE}~\cite{Johnson:2003}. At low density, there is one well-known exception, which corresponds to an enhanced backscattering~\cite{Kuga:1984,vanAlbada:1985,Wolf:1985}. The diffusion coefficient is thus another, important macroscopic quantity, which is governed by the atomic density, and which allows describing the diffuse light. In three-dimensional space it reads~\cite{footnote_diff_coeff}
\begin{equation}\label{eq.D}
D_0 = \frac{v_E \ell_\scat}{3} = \frac{\ell_\scat^2}{3\tau_\tr} \, ,
\end{equation}
where $v_E = \ell_\scat/\tau_\tr$ is the energy transport velocity inside the medium and $\tau_\tr$ the transport time~\cite{Lagendijk:1996,Rossum:1999}. Here, we considered only isotropic scattering such that the transport length equals the mean-free path. For near-resonant light, a remarkable property of cold atomic vapor is that $\tau_\tr = \tau_\at$, the lifetime of the excited state, independently of the detuning~\cite{Lagendijk:1996,Labeyrie:2003}.

In a nonabsorbing passive medium like an atomic vapor probed by a weak intensity laser, the attenuation of the coherent propagating wave is only due to elastic scattering. The mean-free path and the extinction length are thus equal, which leads to the relation
\begin{equation}
\frac{k^3}{6\pi} |\alpha|^2 = \Imag(\alpha)\; ,
\end{equation}
closely related to the so-called optical theorem~\cite{Newton:1976,Lagendijk:1996}. There is thus a close link between the diffusion coefficient, or the mean-free path, governing the transport of the diffuse light, and the susceptibility, governing the transmission of the coherent light.

\subsection{At high atomic density}

When the typical distance between atoms becomes on the order of the wavelength, i.e., at large density $\rho k^{-3} \gtrsim 1$, several nontrivial effects must be taken into account to describe the coherent propagation and the transport of the diffuse light.

First, the relation between the susceptibility and the polarizability has to be modified to include the Lorentz local field~\cite{BornWolf},
\begin{equation}\label{eq.LL}
\chi(\omega) = \frac{\rho \alpha(\omega)}{1-\rho \alpha(\omega)/3} \, .
\end{equation}
This leads to the famous Lorentz-Lorenz shift~\cite{Friedberg:1973},
\begin{equation}
\Delta_\LL = -\pi \rho k^{-3} \Gamma \; ,
\end{equation}
which depends on the density. Here $\Gamma=\tau_\at^{-1}$ is the linewidth of the transition. Although the derivation of Eq.~(\ref{eq.LL}) is based on mean-field arguments and uncontrolled approximations, it yields correct results in many situations~\cite{Lagendijk:1997}. With hot atomic vapors, a density-dependant shift of the resonance line has indeed been observed~\cite{Maki:1991,Keaveney:2012}. With cold atoms, however, recent results suggest that correlations between scatterers at high density are not negligible and the mean-field approximation breaks down~\cite{Javanainen:2016,Jennewein:2016}. The difference with the hot vapor case is probably due to the absence of inhomogeneous broadening, which may suppress the correlations for hot atoms~\cite{Javanainen:2014,Jenkins:2016}. This argument is also consistent with the recent experiment reported in Ref.~\cite{Bromley:2016}.

Closely related to the Lorentz-Lorenz shift is the `cooperative' Lamb shift~\cite{Friedberg:1973,Scully:2009b,Roehlsberger:2010,Keaveney:2012,Meir:2014,Javanainen:2014,Jenkins:2016,Roof:2016}. It is also a density effect but it is influenced by the finite size and the geometry of the medium. We will discuss this effect in more detail in Section~\ref{sec.cooperativity}.
% Actually, in some (or most?) situations, it can be seen as the result of the susceptibility of Eq.~(\ref{eq.LL}) combined with boundary effects, and a standard-optics calculation (wave propagation with refraction, diffraction et partial reflections at the boundaries) yields the correct result~\cite{Javanainen:2016}. For some favorable geometries, like a pencil-shape, a large shift can even appear at low density, without any Lorentz-Lorenz shift, as shown by the recent experiment in the group of Mark Havey~\cite{Roof:2016}. It would be interesting to check if a standard-optics calculation would also explain the result for this special geometry, as was done in~\cite{Jennewein:2016} (in this later case the standard-optics calculation cannot explain the data, because of the high density).

On the mesoscopic side, the diffusion coefficient should also be corrected at high density, this is the so-called ``weak localization'' (WL) correction~\cite{AkkermansMontambaux},
\begin{equation}\label{eq.WL}
D \sim D_0 \left( 1 - \frac{1}{(k \ell_\scat)^2}\right) \;.
\end{equation}
Note that short-path diagrams not captured by the diffusion approximation are expected to yield supplementary $1/k\ell_\scat$ corrections to the transport~\cite{Kirkpatrick:1986,Eckert:2012}. Weak localization corresponds to a slowing down of diffusion due to the constructive interference between reversed-paths of closed loops, which increases the probability of returning to the original point. The parameter $1/(k \ell_\scat)$ quantifies the amount of `disorder'. On resonance, since $\sigma_\scat = \sigma_0=6\pi/k^2$, $1/(k \ell_\scat) \sim 6\pi \rho k^{-3}$.

Weak localization is thought to be a precursor of Anderson (or strong) localization~\cite{Anderson:1958,Lagendijk:2009}, in particular in the self-consistent theory of localization~\cite{Vollhardt:1992}. Anderson localization is expected when $k \ell_\scat \sim 1$ (Ioffe-Regel criterion~\cite{Ioffe:1960}) and corresponds to the complete absence of diffusion. Although weak localization has been observed for light in different systems, including cold atoms using the coherent back-scattering effect~\cite{Labeyrie:1999,Bidel:2002,Kaiser:2005,Labeyrie:2008}, Anderson localization of light in 3D has still not been observed in any experiment~\cite{Beek:2012,Sperling:2016}, and cold atoms may provide a route towards this goal~\cite{Skipetrov:2016b}.

Beyond transport properties, other subtle effects appear in the correlations and fluctuations of scattered light, such as universal conductance fluctuations~\cite{Rossum:1999,AkkermansMontambaux}. These correlations come from common scatterers involved in different paths. If the so-called $C_1$ correlation is short-range and corresponds to the standard speckle grain of traditional statistical optics~\cite{goodman}, the long-range $C_2$ and $C_3$ correlations are highly nontrivial effects, which become nonnegligible when the Thouless conductance $g$ is small, i.e., near the strong localization regime. These effects have not been studied yet using multiple scattering of light in cold atoms.

\section{Collective effects that depend on the optical depth}

As soon as propagation effects are important, the macroscopic local quantities (susceptibility, diffusion coefficient) are not sufficient to describe experimental observables, but should be combined with the size of the medium. The relevant quantity then turns out to be the optical depth.

\subsection{Coherent and diffuse transmission}\label{sec.diffusion}

The complex susceptibility of an atomic medium is measured by the attenuation or the dephasing of a probe wave crossing the sample. It is then obvious that the finite size $L$ of the medium enters the problem. The phase shift induced by the sample will be
\begin{equation}
\varphi = [\Real(n)-1] kL = \Real(\chi)k L/2 \, ,
\end{equation}
and the transmitted intensity will be $T = \exp[-\Imag(\chi)kL]$. The above relation is called the Beer's or Beer-Lambert law. The argument of the exponential is called the optical depth (or thickness) $b$, i.e.,
\begin{equation}
b = -\ln (T) = \Imag(\chi)kL = \rho \sigma_\scat L = L/\ell_\scat \, .
\end{equation}
Although this seems completely trivial, it is interesting to note that a consequence of this exponential attenuation is the nonlinear evolution of the total fluorescence (or `off-axis scattering') with the atom number. The total fluorescence $F$ corresponds indeed to what is not transmitted in the incident mode. We thus get $F \propto 1-e^{-b}$,
which is linear only for low $b$ and saturates for high $b$. This saturation can be intuitively understood as a `shadow effect': atoms at the back of the sample do not radiate because they are not illuminated by the incoming laser. For the same reason, if we sweep the detuning of the probe beam and measure the fluorescence spectrum, it will be a Lorentzian of width $\Gamma$ at low optical thickness only, and deviate from a Lorentzian line shape at larger optical thickness, even though the scattering cross-section for single-atoms $\sigma_\scat = \sigma_0 /(1+4\Delta^2/\Gamma^2)$ is still Lorentzian. More precisely, it induces an effective broadening and a strong directional dependence of the fluorescence, as shown experimentally in~\cite{Labeyrie:2004}. As a consequence, such behaviors are not unique to cooperative effects, even if they are also consistent with a full coupled-dipole model including those effects~\cite{Pellegrino:2014,Sutherland:2016}. The shadow effect can also explain the reduction of the radiation pressure force reported in~\cite{Bienaime:2010,Chabe:2014}, first interpreted as a signature of cooperativity~\cite{Bienaime:2010}. The different collective effects contributing to a modification of the radiation pressure force will be discussed in detail in a forthcoming publication~\cite{Bachelard:tobepublished}.

\begin{figure*}[t]
\centering
\includegraphics{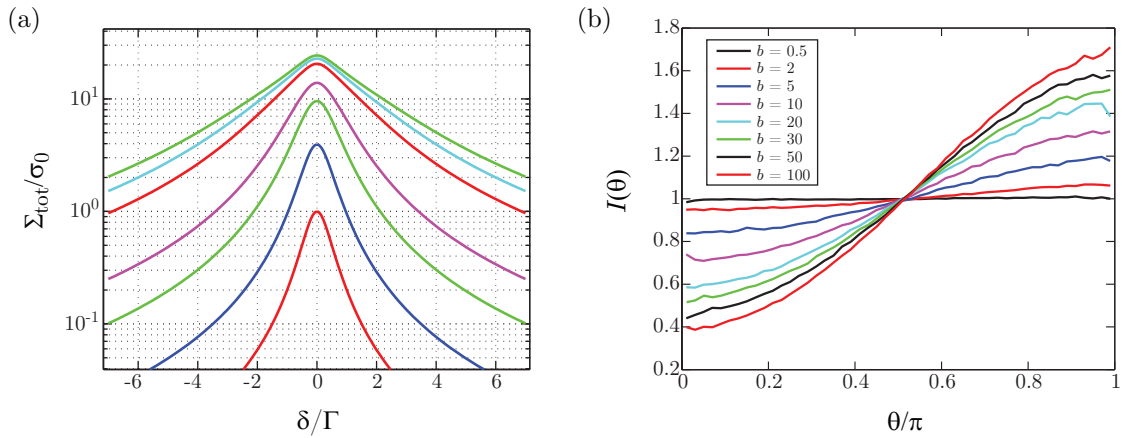}
\caption{Collective effects in the fluorescence, which can be explained by Beer-Lambert law or by multiple scattering. (a) Total fluorescence as a function of the detuning for different atom numbers ($N=1, 5, 20, 50, 500, 325, 450$ from the bottom to the top) and a Gaussian cloud of rms transverse size $R = 0.6\lambda$~\cite{Pellegrino:2014}. Broadening and saturation effects are well visible. (b) Emission diagram of a spherical Gaussian cloud illuminated by a plane wave, computed from random walk simulations, for several optical depths from 0.5 to 100. The emission diagram becomes more backward directed as the optical depth increases.}
\label{fig.fluo}
\end{figure*}

To illustrate the broadening and saturation effects induced by Beer's law, we show in Fig.~\ref{fig.fluo}(a) the behavior of the fluorescence of an atomic cloud as a function of the detuning and the optical depth.
Considering a Gaussian atomic density distribution profile, the total scattering cross-section of the cloud according to Beer-Lambert law, reads~\cite{Chabe:2014,Bachelard:tobepublished}
\begin{equation}
\Sigma_\scat = N \sigma_\scat \times \frac{\Ein(b)}{b} \, ,
\end{equation}
where Ein is the integer function~\cite{Wolfram:Ein}
\begin{equation}
\Ein(b) = \int_0^b \frac{1-e^{-x}}{x} \, dx = b \left[ 1 + \sum_{n=1}^\infty \frac{(-b)^n}{(n+1)(n+1)!} \right] \,.
\end{equation}
Here, $b=\sqrt{2\pi} \rho_0 \sigma_\scat R_z$ is the optical depth that would be measured using a small beam crossing the cloud through its center, $\rho_0$ is the peak density and $R_z$ the rms radius along the propagation axis. Although Beer's law neglects diffraction and refraction effects inside the sample, which are \textit{a priori} not negligible at high $b$ and small nonzero detuning or for very small sample, it appears that such a simplified model is sufficient to qualitatively explain the results of~\cite{Pellegrino:2014}.

The angular dependence is, however, not possible to compute analytically, and numerical simulations based on a random walk process can be used. For an isotropic cloud illuminated by a plane wave, the anisotropy of the emission diagram results from the multiple scattering of light before escaping the sample~\cite{Bachelard:tobepublished}. We illustrate this anisotropy in Fig.~\ref{fig.fluo}(b). It should be noted that it can lead to nonintuitive behaviors. For instance, if light is detected near the forward direction, the measured fluorescence can \emph{decrease} as the atom number increases or as the detuning is reduced (`self-absorption')~\cite{Labeyrie:2004}. Although a simple random walk model completely neglects all wave or coherence effects, it has been shown in Ref.~\cite{Chabe:2014} that, for a large and dilute cloud of two-level systems, the corresponding emission diagram agrees very well with the emission diagram computed from a full coupled-dipole model after averaging over the disorder configurations, except in narrow angular ranges around the forward and backward directions. The forward lobe~\cite{Dicke:1954,Rehler:1971,Scully:2006,Bromley:2016}, absent in an incoherent random walk model, can be explained by diffraction/refraction of light by a continuous index distribution~\cite{Bachelard:2012}. Around backward direction, a narrow cone is visible, which can be explained neither by a random walk model nor by a homogeneous index of refraction. This is a signature of interference in a disordered systems, robust against configuration average and well studied in mesoscopic physics (coherent backscattering)~\cite{Kuga:1984,vanAlbada:1985,Wolf:1985,Labeyrie:1999,Bidel:2002,Kaiser:2005,Labeyrie:2008}. For a Gaussian cloud geometry, analytical expressions for the lobe and cone shapes have been obtained in the double scattering limit~\cite{Rouabah:2014}.

\begin{figure*}[t]
\centering
\includegraphics{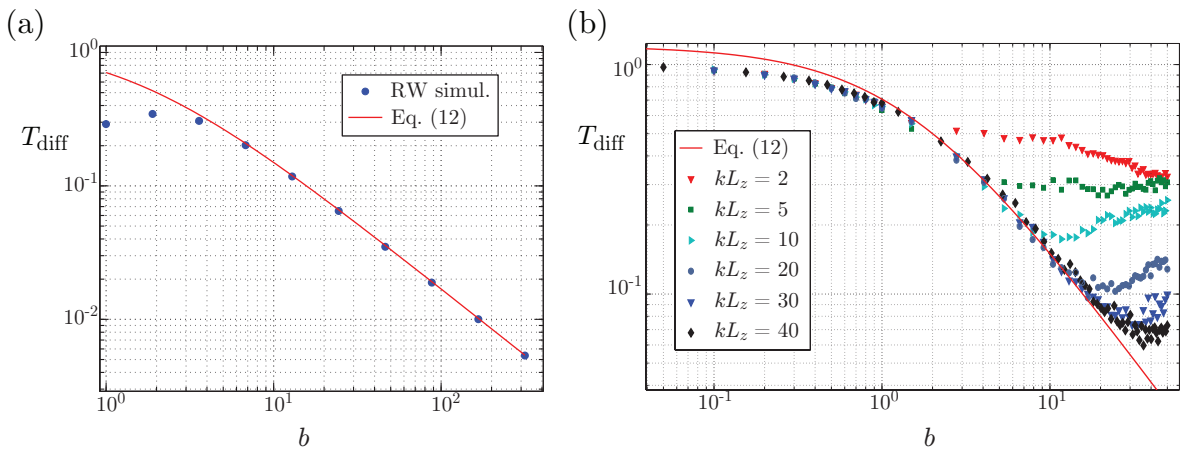}
\caption{Diffuse transmission $T_\diff$ as a function of the optical depth $b$, compared to the asymptotic behavior predicted by the diffusion equation (Ohm's law, solid line) [Eq.~(\ref{eq.Ohm})]. (a) Dots are computed by a random walk simulation. (b) Dots are computed from the coupled-dipole equations in the scalar approximation. Each curve is computed with a given slab depth $kL_z$ and the optical thickness is increased by changing the atom number and thus the density. The driving field is on resonance. At high density, a clear departure from Ohm's law is observed.}
\label{fig.Ohm}
\end{figure*}

Multiple scattering obviously depends on the optical thickness. Although the diffusion coefficient depends on the density, light should first escape the sample before being detected. In steady state, the main consequence of multiple scattering is the complex emission diagram discussed above. Note that for a slab geometry (homogeneous medium of infinite transverse size and finite width $L$), space can be divided into two parts and we can speak of diffuse transmission and diffuse reflection. A well-known result is the asymptotic decrease of the diffuse transmission scaling as $1/L$, which corresponds to Ohm's law for photons [Fig.~\ref{fig.Ohm}],
\begin{equation}\label{eq.Ohm}
T_\diff = \frac{1+ \xi}{b+2 \xi} \, ,
\end{equation}
where $\xi \simeq 0.7104$~\cite{Garcia:1992,Rossum:1999}. This result can be obtained from the diffusion equation or random walk simulations [Fig.~\ref{fig.Ohm}(a)], i.e., a model of light propagation that ignores coherence and interference effects. We have also performed numerical simulations of the diffuse transmission through a slab of atoms, using the coupled-dipole model, in the scalar approximation, where only the far-field term ($\propto 1/r$) of the dipole-dipole interaction is present. We clearly observe a departure from Ohm's law at large density [Fig.~\ref{fig.Ohm}(b)]. The precise origin of the deviation is still under investigation and might be due to cooperativity. Note that the observed deviation corresponds to a relative increase compared to Ohm's law for photons and is thus not consistent with a decrease of the diffusion coefficient due to weak localization. However, it resembles recent observations made with classical scatterers~\cite{Naraghi:2015}, in which the increased transport is attributed to near-field scattering, where evanescent waves open new channels of transmission. In our model though, the near-field terms are neglected, and the physical origin of the increased transmission remains to be explained. Recent numerical studies on the distribution of excitation inside atomic samples of different geometries~\cite{Sutherland:2016} may also be related to our observation.

The temporal dynamics of the diffuse light is also governed by the optical thickness. After some time of illumination, a sudden switch-off of the exciting laser leads to a slow decrease of the fluorescence due to multiple scattering. This `imprisonment of radiation'~\cite{Holstein:1947}, or `radiation trapping'~\cite{Molisch}, has been studied in cold atoms~\cite{Fioretti:1998,Labeyrie:2003}, taking also into account subtle effects like the frequency redistribution induced by the Doppler shift~\cite{Labeyrie:2005,Pierrat:2009} or the multilevel structure~\cite{Baudouin:2013a}. Neglecting those effects, one can easily find the scaling of the radiation trapping time with the optical depth. For a Gaussian random walk in 3D, we have $\left\langle r^2 \right\rangle = 6 D t$. The average number of scattering events for escaping photons is the ratio between the time spent in the system and the scattering time $\tau_\at$,
\begin{equation}
\left\langle N_\scat\right\rangle = \frac{t}{\tau_\at} \sim \frac{\left\langle r^2\right\rangle}{6 D \tau_\at},
\end{equation}
with $D=\ell_\scat^2/(3\tau_\at)$. When $\sqrt{\langle r^2 \rangle} \sim R = b \ell_\scat/2$, the radiation can escape the system, leading to $\left\langle N_\scat\right\rangle \sim b^2/8$.
Radiation trapping times are thus expected to scale as $b^2$, with a precise numerical prefactor that depends on the geometry~\cite{Labeyrie:2003}.

\subsection{Superradiance and subradiance in extended and dilute samples}

The previous effects all depend on the detuning-dependent optical depth,
\begin{equation}
b(\Delta) = \frac{b_0}{1+4\Delta^2/\Gamma^2} \, ,
\end{equation}
where $b_0$ is the on-resonance optical depth. For a cloud of size $R$, and using $\sigma_0\sim 1/k^2$, we have $b_0 \sim N/(kR)^2$, with a numerical prefactor that depends on the geometry.

We now discuss why the super- and subradiant decay rates, measured at large detuning such that attenuation or incoherent multiple scattering are negligible, depend on $b_0$, independently of the detuning, in the case of an extended $R\gg \lambda$ and dilute ($\rho k^{-3} \ll 1$) sample.

%The superradiant decay rate of the Timed-Dicke state~\cite{Scully:2006}, or `phased state', has been computed analytically by many authors~\cite{Arecchi:1970,Rehler:1971,Mazets:2007,Svidzinsky:2008,Svidzinsky:2008b,Courteille:2010,Friedberg:2010,Prasad:2010}, $\Gamma_\supp \propto b_0$, and has been recently observed numerically and experimentally~\cite{Araujo:2016,Roof:2016}. Subradiant decay has also been observed numerically in the coupled-dipole model~\cite{Bienaime:2012} and experimentally~\cite{Guerin:2016}, and it has been shown that the long-time decay is $\Gamma_\sub \propto 1/b_0$.

The physical argument to understand why the cooperativity parameter for super and subradiant decay is $b_0$, is the following. Consider a sample of finite size $\sim R$ radiating in free space. Although the number of modes in free space is infinite, the boundary condition due to the sample surface ($\propto R^2$) sets limitations on the modes that are efficiently coupled to the sample. In particular, the diffraction limit ensures that no mode with a divergence smaller than $\theta~\sim 1/(kR)$ can be emitted from the sample and, in 3D, the total number of modes $M$ efficiently coupled to the sample is related to the sample surface, $M \sim (kR)^2$. It means that if we choose an arbitrary, infinite basis to express the modes in free space, a number $M$ of those modes are enough to describe the radiation pattern of the sample. Then, if the number $N$ of atoms in the sample is larger than the number $M$ of modes, the emission will be cooperative because, in average, $N/M$ atoms emit in the same modes: they are thus coupled to each other via their common coupling to the electromagnetic mode (Fano coupling). We conclude that the `cooperativity parameter' in this problem is the ratio $N/M$, which turns out to be the on-resonance optical depth $b_0$, up to a numerical prefactor. This argument is consistent with the Dicke limit $R\ll\lambda$, for which the only possible outgoing mode is a spherical wave, i.e., $M=1$, and the superradiant enhancement factor is $N$. This reasoning was given in brief in~\cite{Akkermans:2008,Bienaime:2012,Guerin:2016} and is consistent with the classification of superradiance given in~\cite{Longo:2016}.

Let us note that super- and subradiance for $N=2$ has a marked difference with the case of $N\gg2$. First, in order to obtain large dipole-dipole coupling of two atoms in free space, these atoms need to be at a distance comparable to the optical wavelength. When an atomic pair is separated by many wavelengths, interferences between light emitted by these atoms can still occur~\cite{Grangier:1985}. However a modification of the atomic lifetime requires the interatomic distance to be smaller than or comparable to the wavelength~\cite{DeVoe:1996,Barnes:2005}. For small atomic distances, near-field dipole-dipole coupling, scaling as $1/r^3$, becomes dominant, while super- and subradiance obtained for many atoms in the dilute limit is relying on the far-field dipole coupling, scaling as $1/r$. This long-range behaviour of the far-field dipole-dipole coupling is essential for the collective scaling of the Dicke subradiance observed in~\cite{Guerin:2016}. Let us also note that the competition between the near and far-field dipole-dipole coupling has been discussed in~\cite{Friedberg:1974,Milonni:1974} and has been called `van~der~Waals dephasing' in~\cite{Gross:1982}. It is interesting to note that while for $N=2$, the eigenstates of the near-field part of the interaction Hamiltonian $H_\NF$ are also eigenstates of the far-field interaction $H_\FF$ (both being proportional to the same $\sigma_1$ Pauli matrix), for $N>2$, $H_\NF$ and $H_\FF$ are not proportional if the distances between all atoms are not equivalent (which can be obtained in the particular case of $3$ atoms equally spaced on a ring or four atoms equally spaced on a sphere). In this case $H_\NF$ and $H_\FF$ do not commute and have different eigenvectors (see Sec.~4 of~\cite{Gross:1982}). This particular situation does not prevent the total Hamiltonian to have eigenvalues corresponding to long-lived (subradiant) and short-lived (superradiant) modes. However the symmetries of the eigenstates (and potentially their sensitivity to perturbations) will differ from those of $H_\FF$.

\begin{figure*}[t]
\centering
\includegraphics{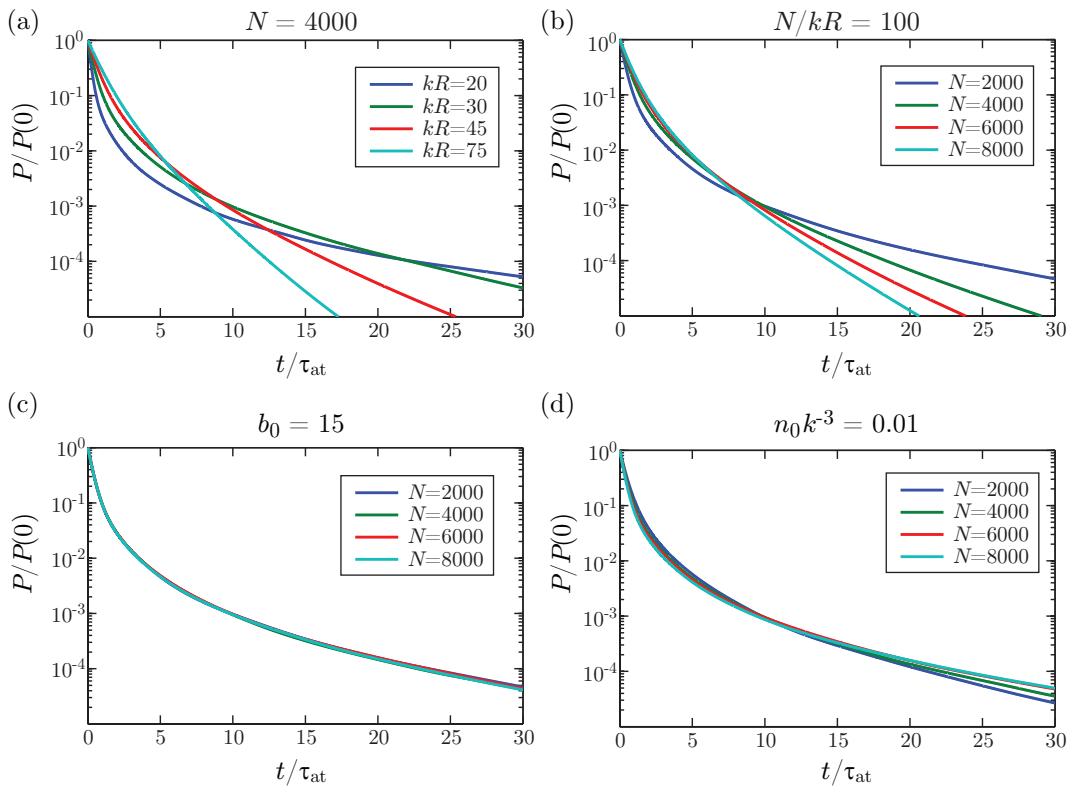}
\caption{Decay of the total scattered power after the switch-off of the driving laser at $t=0$ computed from the coupled-dipole model in the scalar approximation, as in~\cite{Guerin:2016}, averaged over only two configurations. The time axis is normalized to the lifetime of the excited state $\tau_\at$ and the vertical axis to the steady-state scattered power. The driving laser has a large detuning, $\Delta=50\Gamma$, and has been applied for a duration of $30\tau_\at$ so that the system reaches its steady state. In panel (a), the atom number $N=4000$ is kept constant and the size is varied, $b_0=3N/(kR)^2$ is varied from $30$ to $2.3$. In panel (b), both $N$ and $R$ are varied such that $N/kR$ is kept constant, $b_0$ is varied from $15$ to $3.75$. In panel (c), $N$ and $R$ are varied such that $b_0=15$, and in panel (d) $N$ and $R$ are varied such that the density is kept constant, $n_0 k^{-3} = 0.01$, and $b_0$ is varied from $11$ to $17.5$.}
\label{fig.decay}
\end{figure*}

As explained from the beginning, it is not surprising that the size of the sample is an important parameter to take into account, since the superradiant and subradiant decays are transient phenomena, which means that light must escape the sample. However, from a mathematical point of view, when looking at the coupled-dipole model~\cite{Ruostekoski:1997,Javanainen:1999,Courteille:2010,Svidzinsky:2010,Bienaime:2011}, it is not obvious which combination of the parameters yields to correct scaling laws for the different features that can be observed. In this model, once the geometry of the cloud is fixed (and spherical), there are three independent parameters: the atom number $N$, the size $R$ and the detuning of the driving laser. In the limit of a very large detuning, after a long enough illumination, the system reaches a steady state given by the `timed-Dicke state'~\cite{Scully:2006,Courteille:2010,Bienaime:2011}. Once the laser is switched off, the decay of this state could be governed by any combination of $N$ and $R$. In particular, it might be considered more intuitive that the density $\rho \sim  N/R^3$ is the important parameter, in particular because the dipole-dipole coupling terms, which decay as $1/(kr)$ with the interparticle distance $r$, become very small as the density decreases. But even if the coupling terms are small, each atom is coupled to $N-1$ other atoms, and $N$ can be huge, which, somehow, balances the low density and makes $b_0 \sim N/(kR)^2$ to be the scaling parameter. Note that the $1/r$ term in the dipole-dipole interaction bears all the relevant ingredient for  `long-range' interaction, which, as in gravitational or Coulomb interactions, yields to well-known size-dependent effects~\cite{Dauxois:2010}. This parameter also appears in the statistical properties of the eigenvalues of non-Hermitian Euclidean random matrices such as the one involved in the couple-dipole model~\cite{Skipetrov:2011,Goetschy:2011b} and the early (superradiant) decay of the timed-Dicke state~\cite{Scully:2006}, or `phased state', has been computed analytically by many authors, $\Gamma_\supp \propto b_0$~\cite{Arecchi:1970,Rehler:1971,Mazets:2007,Svidzinsky:2008,Svidzinsky:2008b,Courteille:2010,Friedberg:2010,Prasad:2010}. Numerical and experimental evidence have been given in~\cite{Guerin:2016,Bienaime:2012} that the long time decay is also governed by $b_0$, $\Gamma_\sub \propto 1/b_0$. Here, we show in Fig.~\ref{fig.decay} that not only the short and long time limits are both depending on $b_0$, but the full decay curve is in fact only dependent on $b_0$. An analytical function for describing this decay curve remains to be found. We have checked that this is neither a power law, nor a stretched exponential.

\section{The mean-field approach of traditional optics and the independent scattering approximation of mesoscopy}

It was pointed out in recent theoretical~\cite{Javanainen:2016} and experimental studies~\cite{Jennewein:2016} that Eq.~(\ref{eq.LL}), which includes Lorentz local field correction, is inaccurate for cold atoms at high density.  Indeed, another term, usually not considered, has the same order of magnitude. Following ~\cite{Jennewein:2016}, the more precise expression is given by
\begin{equation}\label{eq.LL_modified}
\chi(\omega) = \frac{\rho \alpha(\omega)}{1-\rho \alpha(\omega)(1/3+\beta(\omega))} \, ,
\end{equation}
where $\beta(\omega) \sim \alpha(\omega)k^3$ and specifically depends on the geometry. This term is neglected in the mean-field approach leading to Eq.~(\ref{eq.LL}) even though on resonance, $\alpha(\omega) \sim k^{-3}$, and thus $\beta(\omega)$ has the same order of magnitude than the $1/3$ corresponding to the density-dependent Lorentz local-field correction. To our knowledge, this result has been first obtained by Saunders and Bullough~\cite{Saunders:1973a,Saunders:1973b} (see Eq.~(22) of \cite{Saunders:1973b}) and has been later rediscovered several times~\cite{Morice:1995,Ruostekoski:1997}. One hypothesis for the success of Eq.~(\ref{eq.LL}) could be that the mean-field result is recovered if some mechanism breaks the correlation between scatterers, as it could be the case, e.g., for hot vapors with the Doppler broadening~\cite{Javanainen:2014,Jenkins:2016}.

At this point, it is interesting to point out the close link between Eq.~(\ref{eq.LL_modified}) and the weak localization correction of the diffusion coefficient (Eq.~\ref{eq.WL}). First, we note that Eq.~(\ref{eq.LL_modified}) can be rewritten
\begin{equation}
\chi(\omega) = \frac{\rho \alpha(\omega)}{1-\rho \alpha(\omega)/3 -\rho \alpha(\omega)\beta(\omega)} \, ,
\end{equation}
which isolates a correcting factor $\rho \alpha(\omega)\beta(\omega)$ in the usual Lorentz-Lorenz formula (\ref{eq.LL}). Then, if $\beta(\omega) \sim \alpha(\omega)k^3$, this correcting factor is $\sim \rho  k^3 \alpha(\omega)^2$, which is almost identical to
\begin{equation}
\frac{1}{k\ell_\scat} = \frac{\rho \sigma_\scat}{k} = \frac{\rho k^3 |\alpha(\omega)|^2}{6\pi} \;,
\end{equation}
the disorder parameter used in mesoscopic optics. Given the close link between the diffusion coefficient and the susceptibility, there is a strong similitude between the correction of Eq.~(\ref{eq.LL_modified}) and the weak localization correction to the diffusion coefficient Eq.~(\ref{eq.WL}).

This similitude is not surprising. The mean field approach of traditional optics neglects correlation between scatterers. When they exist, these correlations are due to the nonnegligible probability that a photon goes back to a previous scatterer, a process sometimes called recurrent scattering. In the mesoscopy community, neglecting this recurrent scattering and the subsequent correlations is called the `independent scattering approximation' (ISA)~\cite{Lagendijk:1996,Rossum:1999,AkkermansMontambaux}. Weak localization effects are typical signatures of `dependent scattering'.
The link between the ISA and the Lorentz-Lorenz local field theory has been already briefly discussed in Ref.~\cite{Lagendijk:1997} but is often not discussed in more recent work.

%, with a conclusion that apparently disagrees with the ones of~\cite{Javanainen:2016,Jennewein:2016}, since Ref.~\cite{Lagendijk:1997} concludes that the Lorentz-lorenz formula turns out to be correct.

% Maybe there should also be a link with the theory of Homogenization of Maxwell's equations, which also has some revival in the context of metamaterials, although they mainly deal with periodic dielectrics (with almost real susceptibility). The proposals and experiments on photonic band gaps in ordered cold-atom systems are going in this direction.

\section{What should be called cooperative?}\label{sec.cooperativity}

Based on the previous discussion, we consider that the term `cooperative scattering' should not be kept for effects related to recurrent scattering only, because it is already called `dependent scattering' in the mesoscopy community, and several such effects have well-established names, like weak or strong localization. Moreover, effects related to Dicke superradiance and subradiance can appear at low density, as shown in Fig.~\ref{fig.decay} and in~\cite{Araujo:2016,Roof:2016,Guerin:2016}, without recurrent scattering, and have been called `cooperative' for a very long time (see, e.g., Refs.~\cite{Arecchi:1970,Rehler:1971,Haake:1972,Stroud:1972,Saunders:1973b,Friedberg:1974,Milonni:1974,Bonifacio:1975}). Changing the semantics without a clear distinction of previously studied effects in different communities is therefore probably not recommendable.

However, we agree that it is questionable to call collective effects `cooperative' if these effects can also be explained by conceptually more simple models, like wave propagation in a medium characterized by its susceptibility or standard multiple scattering.

In addition to the examples already mentioned~\cite{Bienaime:2010,Pellegrino:2014}, the coherent forward emission~\cite{Dicke:1954,Rehler:1971,Scully:2006,Bromley:2016} is certainly such an effect. Even if using a single-photon \textit{Gedankenexperiment} requires a quantum formalism to properly demonstrate that the absorbed photon will be preferentially emitted in the forward direction~\cite{Scully:2006}, this effect is a simple consequence of an $N$-wave interference, like in a multiple slit experiment, and it is well-known that slit experiments produce interference patterns even with single particles~\cite{Tonomura:1989}. The interference pattern from two atoms radiating a single photon has also been observed in~\cite{Grangier:1985,Eichmann:1993}. The analogy between $N$-slit interference and the directionality character of superradiance has been further developed, including higher-order correlations, in Ref.~\cite{Oppel:2014}.
In the CW regime, considering the phase-matching in the forward direction is also a way to understand the index of refraction of a polarizable medium. Thus, the resulting forward lobe is actually the diffraction/refraction pattern of the sample, like a Mie scatterer~\cite{Bachelard:2012}.

The `cooperative' Lamb shift is another questionable example.
%In the case of a mesoscopic ensemble containing a small number of particles, like in the experiment of Ref.~\cite{Meir:2014}, the cooperative nature of the shift is not ambiguous. However,
% On pourrait plutot dire que le passage de N=2, N>2 à N>>2 est certainement intéressant pour comprendre la construction ab initio d'une grandeur collective.
In the case of macroscopic samples, for which a susceptibility can be defined,
 %(using $\chi = \rho \alpha$ at low density~\cite{Roof:2016} or Eq.~(\ref{eq.LL}) at high density if the mean-field approach turns out to be correct~\cite{Keaveney:2012})
Javanainen and Ruostekoski have shown~\cite{Javanainen:2016}, at least for the slab geometry that applies to the experiment of Ref.~\cite{Keaveney:2012}, that the `cooperative' Lamb shift was the result of `standard optics', i.e., wave propagation in a dielectric medium following the Helmholtz equation. It would be interesting to know, using wave propagation simulations, if this also holds for the cigar-shaped sample used in Ref.~\cite{Roof:2016} or any other geometry~\cite{Manassah:2012}. For example, one of the geometries studied by Manassah~\cite{Manassah:2012} consists of a density-modulated slab of atoms and a `giant cooperative Lamb shift' was predicted when the density is modulated at the Bragg condition~\cite{Manassah:2010}. Such an experiment has already been performed~\cite{Schilke:2011}, and very asymmetric and shifted Bragg-reflection spectra have been indeed reported (see, e.g., Fig.~2 of~\cite{Schilke:2011}). These feature can be explained by the coupled-dipole model~\cite{Samoylova:2014}, but they can also be very well described by a standard-optics wave-propagation equation based on a periodic susceptibility (in that case using transfer-matrices to exploit the periodicity). Thus, it seems that in many (or most?) situations, the `cooperative Lamb shift' is in fact a collective shift related to the shape and finite size of the medium, inducing refraction, reflection, lensing and waveguide effects~\cite{Roof:2015}, etc., which can be simulated by wave-propagation simulations.

The problem is that such wave-propagation simulations are not simple to perform. In the recent years, the coupled-dipole model has been widely used~\cite{Ruostekoski:1997,Svidzinsky:2008b,Courteille:2010,Svidzinsky:2010,Bienaime:2011,Chomaz:2012,Miroshnychenko:2013,Meir:2014,Pellegrino:2014,Feng:2014,Guerin:2016,Sutherland:2016,Araujo:2016}. It is computationally limited to a few thousand atoms, but it is otherwise very simple to use. Another strength is its completeness: it includes all collective effects we have discussed so far. But this is also a drawback: it is sometimes hard to give simple interpretation of the numerical results and, when comparing to experimental results, a good agreement does not help identifying the relevant physical ingredients of the experiment. As a consequence, it would be wise, although tedious, to systematically compare the results with simulations based on a standard-optics calculation (wave propagation in a dielectric) or with a standard multiple-scattering computation (random walk), to check if the observations do not have simple explanations. This methodology has been used in~\cite{Guerin:2016,Jennewein:2016}.

It seems to us that less ambiguous signatures of cooperativity can be found in the transient response of the system, as initially envisioned by Dicke for the superradiant emission of a fully inverted system~\cite{Dicke:1954,Feld:1980,Gross:1982} and its subradiant counterpart~\cite{Pavolini:1985}. In the single-photon or linear-optics regime, superradiant decay rates have been observed recently in the forward direction~\cite{Roof:2016} and also off-axis~\cite{Araujo:2016}.

In the forward-direction case, the measured light is the diffracted/refracted light by the cloud. Its dynamics is thus related to the dynamics of the refractive index. The imaginary part of the refractive index is also what governs the transmission, and its dynamics at the switch-on or -off of the incident laser is what gives rise to optical precursors~\cite{Jeong:2006,Chen:2010} or `flashs'~\cite{Chalony:2011,Kwong:2014,Kwong:2015}. It has been shown experimentally and theoretically, using only Beer-Lambert law including the dephasing, that dynamics faster than $\Gamma$ can appear at large on-resonance optical thickness~\cite{Kwong:2015}. This is somehow the counterpart of the spectral broadening induced by Beer's law at high $b_0$.

Since this broadening is also present for the light scattered off-axis (see Fig.~\ref{fig.fluo}a), one can wonder if similar arguments also explains off-axis superradiance. Off axis, though, the field arriving on the detector emitted by different atoms have random phases (no phase-matching). There is thus no straightforward calculation similar to the one of~\cite{Kwong:2015}, at least to our knowledge, that could relate the spectral broadening of the system response to the temporal dynamics observed on the fluorescence. This is, we think, the nontrivial feature of cooperativity and superradiance in the linear regime that this accelerated decay is still preserved despite the absence of any obvious phase-matching condition. Thus, the most efficient way to understand this result is to invoke the collective (cooperative) modes of the atomic excitation in the cloud, some of them decaying fast (superradiant modes), and some decaying slowly (subradiant modes). In other words, we trace over the photon degrees of freedom and look at the collective (cooperative) atomic behavior. The reverse approach, looking at the light, scattered by all atoms and interfering, is certainly possible and might provide an alternative view on the light-matter interaction for many atoms and at low intensities.% This is, probably, the situation where it should be called `cooperative scattering'.

Subradiant decay, as the counterpart of superradiance, is also a good signature of cooperativity. However, a particular care is necessary to distinguish between different effects that could lead to a slow decay of the scattered light. In a dense sample with sharp boundaries, for example, long-lived cavity-like modes (or `polaritonic modes'~\cite{Schilder:2016}) may exist, and may also give rise to high-Q Mie-like resonances~\cite{Bachelard:2012}. If the optical thickness $b(\Delta)$ is large (near resonance), multiple scattering may trap the light in the sample for a long time (radiation trapping, see Sec.~\ref{sec.diffusion} and ref. \cite{Labeyrie:2003}), and even strong localization might need to be considered~\cite{Skipetrov:2016}.
In that case, the qualitative difference between those various effects can be seen in their different scalings with the experimental parameters. The radiation trapping time scales as $b(\Delta)^2$ whereas the subradiant time scales as $b_0$~\cite{Guerin:2016}, independently of the detuning. Here again, Mie resonances or radiation trapping are well understood by looking at the light, whereas subradiance is well understood by looking at the collective modes of the atomic excitation. If we want to explain subradiance with a physical picture using light, we can say that coherence and interference effect modifies multiple scattering such that some `lucky photons' are trapped in the sample, even at large detuning. If we neglect coherence and interference effects (diffusion or random walk model), we find, on the contrary, that multiple scattering vanishes far from resonance. The interplay between this incoherent radiation trapping and subradiance for intermediate detuning, and the way to distinguish between them, is the subject of our current investigation.

\section{Conclusion}

In this article, we have tried to discuss the meaning of `cooperative scattering' from a broad perspective, by studying various collective effects in atom-light interaction. We conclude that what should be called cooperative is often mainly a question of perspective. When we describe collective effects by looking at the atoms, with a microscopic modeling (based on Dicke states, coupled dipoles, etc.), most features appear to be of cooperative nature. Nevertheless, one can also understand many of these effects using the point of view of light, undergoing multiple scattering or propagating through a macroscopic sample, without using the concept of cooperativity at all.

The temporal decay of the light emitted off-axis seems to be the best example so far of a situation where a cooperative-scattering approach is the most efficient, numerically as well as conceptually, to describe the physics. We have shown that, at large detuning, the full decay curve only depends on the on-resonance optical thickness of the sample (and of its precise shape), but an analytical description of this result is still missing.

Finally, we emphasize that most, if not all, of the effects that we discussed in this article, are already known, but not always in the same fields. Superradiance and subradiance are hardly known in the mesoscopy community, and the same can be said for effects related to multiple scattering with the quantum-optics community. Establishing a bridge between those communities and their concepts seems very important to us. In particular, we have pointed out the similarity between the recurrent-scattering corrections to the susceptibility recently discussed in the quantum-optics community~\cite{Jennewein:2016,Javanainen:2016} and the weak-localization corrections, which are well-known in the mesoscopy community. We have also presented new results on the deviation from Ohm's law predicted by the coupled-dipole model. The physical origin of this deviation remains to be understood, but this is an attempt to relate cooperativity to \emph{transport properties}~\cite{Gero:2006}. This connection may be fruitful for the question of Anderson localization of light in 3D~\cite{Skipetrov:2016,Skipetrov:2016b} or in relation with other topics, like superconductivity~\cite{Baskaran:2012} or light-harvesting systems~\cite{Monshouwer:1997,Celardo:2014}.

%To conclude, we hope to have clarified something, and maybe we have put some bridge between different communities or concepts.
%The two new results are the deviations from Ohm's law, to be explained, and the full decay curve depending on $b_0$, to be understood by mathematicians (Akkermans, Skipetrov).

%\begin{acknowledgement}
We thank Antoine Browaeys, Yvan Sortais and Mark Havey for fruitful discussions at the PQE conference and Joachim von~Zanthier for his invitation.
This work was supported by the Agence National pour la Recherche under grant No. ANR-14-CE26-0032 (project LOVE). M.T.R. was supported by an Averro\`es exchange program.
%\end{acknowledgement}

%\bibliographystyle{D:/RECHERCHE/MesPublis/MaBiblio/prsty_no_etal}
%\bibliography{D:/RECHERCHE/MesPublis/MaBiblio/AllMyBiblio}

\end{document}